\documentclass[aps,prl,showpacs,twocolumn,floatfix,superscriptaddress]{revtex4-1}

\usepackage{amsmath,amssymb}
\usepackage{graphicx}
\usepackage{dcolumn}
\usepackage{bm}
\usepackage[usenames]{color}
\usepackage[breaklinks=true,colorlinks=true,linkcolor=blue,urlcolor=blue,citecolor=blue]{hyperref}
\usepackage{ulem}
\usepackage{amsmath}
\usepackage{ulem}

\begin{document}

\date{\today}

\title{Structural dynamics and optimal transport of an active polymer}

\author{Hamidreza Khalilian}
\email{khalilian@ipm.ir}
\affiliation{School of Nano sciences, Institute for Research in Fundamental Sciences (IPM), 19395-5531, Tehran, Iran.}

\author{Fernando Peruani}
\email{Fernando.Peruani@cyu.fr}
\affiliation{Laboratoire de Physique Th{\'e}orique et Mod{\'e}lisation, UMR 8089, CY Cergy Paris Universit{\'e}, 95302 Cergy-Pontoise, France.}

\author{Jalal Sarabadani}
\email{jalal@ipm.ir}
\affiliation{School of Nano sciences, Institute for Research in Fundamental Sciences (IPM), 19395-5531, Tehran, Iran.}


\begin{abstract}
We study the spontaneous configuration transitions of an active semi-flexible polymer 
between  {\it spiral} and {\it non-spiral} states, and show that 
the configuration dynamics is fully described by a {\it subcritical pitchfork} bifurcation.  
Exploiting the fact that an active polymer barely moves in  {\it spiral} states and exhibits net displacements in {\it non-spiral} states, we theoretically  prove that the motion of the active polymer is consistent with a {\it run-and-tumble}-like dynamics. 
Moreover, we find that there exists an {\it optimal} self-propelling {\it force} that maximizes the diffusion coefficient.
\end{abstract}

\maketitle


\section{Introduction}

Active polymers have recently become an interesting field of research in soft matter and non-equilibrium statistical physics. They play an important role in biological systems such as 
synthesizes of proteins by ribosomes~\!\cite{albert, zia}, chromatin in eukaryotic cells~\!\cite{chromatin1,chromatin2,chromatin3,chromatin4,chromatin5,chromatin6}, movement of actin filaments by motor proteins in cellular cytoskeletons~\!\cite{cytos1, cytos2, turning6, turning7,Motility_Assay_Book}, microtubule bundles and spools in motility assays~\!\cite{Chaudhuri_Nano_Lett_2018}, mechanical sensing of soft materials~\!\cite{Inoue_Nat_Commu_2015}, directed transport and molecular sorting of microtubules in kinesin-coated nanostructures~\!\cite{Dekker_NanoLett_2005,Maximov_NanoLett_2013,Dekker_Science_2006}, and the coordinated beating of flagella~\!\cite{flagella0,flagella1,flagella2}. 
In all of the above examples, as mechanical tangential forces are generated on the backbone of the elongated objects, these can be considered as self-propelled polymers. Motivated by these applications, many studies have focused, both experimentally~\!\cite{exp1,exp2,exp3,exp4,exp5} and theoretically~\!\cite{theo1,theo2,theo3,theo4,theo5,theo6,theo7,theo8,theo9,PRL,theo10,
baskaran,theo11,turning,theo12,Jabbari_PRE,Jabbari_JCP}, on active polymers.   
Theoretical studies have shown that the diffusion coefficient of  flexible/semi-flexible active polymers is a monotonically increasing function -- linear or quadratic -- of P{\'e}clet number~\!\cite{theo1,theo5,theo6,theo7,theo9,PRL,baskaran,theo11,theo12}. 
However, we note that the active forces that are applied on the polymer are not necessary tangential to it. This can occurs, for instance, due to the presence of an  active and/or viscoelastic bath~\!\cite{ViscoBath} or enzymatic force dipoles~\!\cite{EnzymActive}. While such systems are undoubtedly of great relevance, here our focus is on active polymers driven by tangential forces.

\begin{figure}[tb]
\begin{center}
\includegraphics[width=0.25 \textwidth]{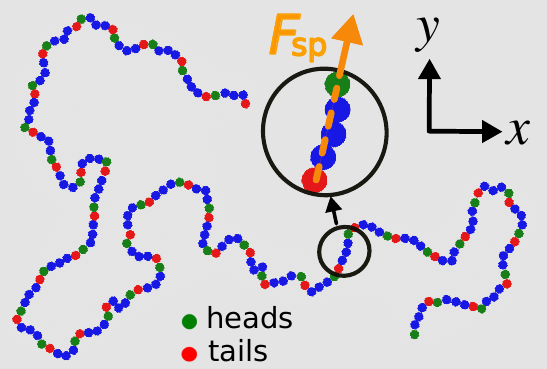}
\end{center}
\caption{Typical snapshot of a semi-flexible polymer with contour length of $N=200$ and persistence length of $\ell_{\textrm{p}}=5$ after equilibration and just before acting the self-propelling force $F_{\textrm{sp}}$ (orange arrow) on each segment. The polymer is divided to $N/\ell_{\textrm{p}}$ segments. The red and green beads represent the tail and head of each segment, respectively. In each segment  $F_{\textrm{sp}}$ acts on the head monomer directing from tail to head (orange dashed line).
} 
\label{fig1}
\end{figure}
\begin{figure}[t]
\begin{center}
\includegraphics[width=0.47 \textwidth]{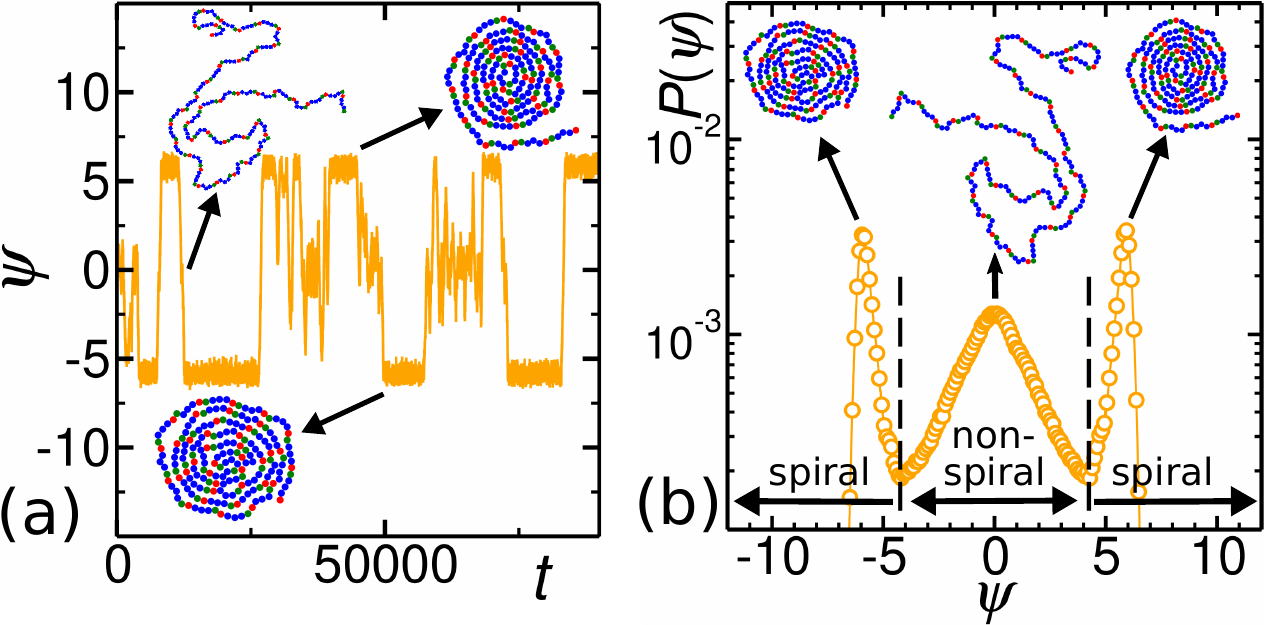}
\end{center}
\caption{(a) The turning number $\psi (t)$ as a function of the time $t$ for fixed values of the polymer contour length of $N=200$, persistence length of $\ell_{\textrm{p}} = 5 $ and the SP force of $F_{\textrm{sp}}=2$. The top and the bottom insets are typical snapshots of the polymer in the {\it spiral} state with $\psi>0$ (CCW) and $\psi<0$ (CW), respectively. In each segment the SP force $F_{\textrm{sp}}$ acts on the green circle directing from the red circle (tail) to the green one (head).
(b) Definition of the {\it spiral} and {\it non-spiral} states: probability distribution function of the turning number $P(\psi)$ for the same values of $F_{\textrm{sp}}$, $N$ and $\ell_{\textrm{p}}$ as of panel (a). The vertical black dashed lines at minima of the $P(\psi)$ (located at the left $\psi_{\textrm{min}}^{\textrm{l}}$ and the right $\psi_{\textrm{min}}^{\textrm{r}}$ minima) separate the two {\it spiral} and {\it non-spiral} states. The polymer with $\psi_{\textrm{min}}^{\textrm{l}} < \psi < \psi_{\textrm{min}}^{\textrm{r}}$ is in the {\it non-spiral} state, and it is in the {\it spiral} state for $\psi < \psi_{\textrm{min}}^{\textrm{l}}$ or $\psi > \psi_{\textrm{min}}^{\textrm{r}}$. The right and left snapshots correspond to the {\it spiral} states with counter-clockwise and clockwise turnings, respectively, and the middle one exhibits a {\it non-spiral} state. See SM for movies.} 
\label{fig2}
\end{figure}

The structural dynamics of an active polymer has been studied in Refs.~\!\citenum{theo4,theo5,theo6,theo9,PRL,theo10,baskaran,turning,theo11,Jabbari_PRE,Jabbari_JCP}. 
The interplay between thermal fluctuations and self-propelling stresses (i.e. activity)~\cite{theo9,baskaran,turning} -- as occurs when a polymer is embedded 
in a gliding assay with motor proteins as the source of activity~\!\cite{Duke_PRL,Bourdieu_PRE,Bourdieu_PRL,Grill_Nano_Lett,Tas_Nano_Lett,Farkas_book,theo10,turning} -- leads to 
transient polymer configurations, classified into two main groups: {\it spiral} and {\it non-spiral}  ones. 
For the latter case~\!\cite{turning} the radius of gyration (RG) is a convex function with respect to the attachment/detachment rates and P{\'e}clet number of the motor proteins, while for a system composed of an active polymer wherein the force is applied to all beads of the polymer~\!\cite{theo9}, the RG is a monotonic decreasing function of the polymer activity [see the Appendix~\!A for more details].

In the present paper, we show that transitions between configurational states of the polymer is consistent with a   {\it subcritical pitchfork} bifurcation. Furthermore, we prove that the diffusion coefficient for the polymer center of mass (CM), $D_{\textrm{CM}}$ 
is, counterintuitively, a non-monotonic function of the self-propelling (SP) force, $F_{\textrm{sp}}$, acting on different segments of the polymer. Moreover, we find that there exist an optimal self-propelling force 
$F_{\textrm{sp}}$ that maximizes $D_{\textrm{CM}}$.


\section{Simulation model}

We consider a bead-spring semi-flexible polymer with $N=200$ monomers in a two-dimensional, squared simulation box of area $L\!\times\!L$ --  where  
$L=500\sigma$ and $\sigma$ the length unit -- with periodic boundary conditions (Fig.~\!\ref{fig1}). The $N$ monomers are disks of radius $\sigma$ and mass $\textrm{M}$.
The consecutive monomers are connected by the finitely extensible nonlinear elastic (FENE) potential $U_{\textrm{FENE}}(r) = -\frac{1}{2}k R^{2}_{0}\ln[1-(r/R_{0})^{2}]$, where $r$ is the center-to-center distance between two connected monomers, $k$ the spring constant, and $R_{0}$ the natural length of the spring.  
Monomers interact via the repulsive Weeks-Chandler-Anderson (WCA)~\!\cite{wca} potential $U_{\textrm{WCA}}(r)= U_{\textrm{LJ}}(r)-U_{\textrm{LJ}}(r_{\textrm{c}})$ if $r \leq r_{\textrm{c}}$ and zero otherwise, where $r_c=2^{1/6}\sigma$ and $r$ are the cut-off radius and the distance between two given monomers, respectively. $U_{\textrm{LJ}}(r) = 4 \epsilon \big[ \big( \sigma / r \big)^{12} - \big( \sigma / r \big)^6 \big]$ is the Lennard-Jones (LJ) potential, with $\epsilon$ as the depth of the potential well. 
The rigidity of the semi-flexible polymer is modeled by the cosine potential $U_{\textrm{bend}} (\alpha_i) = \kappa_{\textrm{b}}[1+\cos\left(\alpha_{i}\right)]$, where $\kappa_{\textrm{b}}$ is a bending constant,  $\alpha_{i}$ represents the angle between two consecutive bond vectors connecting $i$th and $i+1$th, and the $i+1$th and $i+2$th beads, 
whose positive and negative values correspond to the counter-clockwise (CCW) and clockwise (CW) local configurations of the consecutive $i$th and $(i+1)$th bonds, respectively. 
For a non-zero value of the $\kappa_{\textrm{b}}$, the persistence length of the polymer in two dimensions is obtained as $\ell_{\textrm{p}}=2\kappa_{\textrm{b}}/(k_{\textrm{B}}T)$. The $k_{\textrm{B}}T$ is the thermal energy and $k_{\textrm{B}}$ is the Boltzmann constant.
For the sake of (computational) simplicity, the polymer is divided into $N/\ell_\textrm{{p}}$ segments. Each segment -- see Fig.~\!\ref{fig1}-- contains a tail (in red color) and a head (in green color) monomers as the first and last monomers, respectively, and the rest of monomers are in blue color. Importantly, a self-propelling  force $F_{\textrm{sp}}$ acts on the head monomer of each segment and in the direction from tail to head, depicted by the orange arrow and dashed line, respectively in Fig.~\!\ref{fig1}~\!\cite{Amir}. 

Employing  Langevin dynamics (LD), 
the temporal evolution of the $i$th monomer is given by the equation of motion:
\begin{equation}
M\ddot{\vec{r}}_{i} = -\gamma \dot{\vec{r}}_{i} + \vec{F}_{\textrm{sp}}\delta_{\textrm{ih}} - \vec{\nabla}_{r}U_{\textrm{tot}} + \sqrt{2\gamma k_{\textrm{B}}T} \vec{\eta}_{i}(t),
\label{langevin_equ} 
\end{equation}
where $U_{\textrm{tot}}=U_{\textrm{FENE}}+U_{\textrm{bend}}+U_{\textrm{LJ}}$ is the total potential energy and $\gamma$ is the solvent friction coefficient that the $i$th monomer experiences. In the Kronecker delta $\delta_{\textrm{ih}}$ the $\textrm{h}$ stands as an index of the head monomer in each segment. The white noise term follows $\langle \eta_{i}^{m} (t) \rangle  =0$ and $\langle \eta_{i}^{m} (t)\eta_{j}^{l} (t') \rangle  = \delta_{ij}\delta_{ml}\delta (t - t')$, where $m,l\equiv x, y$. The $\textrm{M}$, $\sigma$ and $\epsilon$ are used as the simulation unit scales for mass, length and energy, respectively. The temperature is kept at $k_{\textrm{B}}T=1.2\epsilon$ and the friction coefficient is set to $\gamma=0.7\tau_{0}^{-1}$, in which $\tau_{0}=\sqrt{ \textrm{M} \sigma^{2}/\epsilon}$ is the LJ time unit. In our simulations, we consider $dt=0.001\tau_{0}$ and $t_{\textrm{eq}}=5\times 10^{4}\tau_{0}$ as the time step for the integration of the equations of motion and the equilibration time interval, respectively. We also set $k=30$, $R_{0}=1.5$ and $\kappa_{\textrm{b}}=3$ to have a semi-flexible polymer with persistence length of  $\ell_{\textrm{p}}=5$. 
The size of each bead corresponds to the single-strand DNA Kuhn length of  $\sigma \approx 1.5$nm. The strength of the interaction at room temperature ($T=295$K) is $3.39 \times 10^{-21}$J and the mass of each bead is $936$~\!amu. Thus, the LJ time scale  is 32.1ps. The parameters and variables in the LD simulations are expressed in the LJ units. 
The polymer is initially equilibrated during the time interval $t_{\textrm{eq}}$ with $F_{\textrm{sp}}=0$. Then the SP force is switched on and the main simulations are done for $10^{6}\tau_{0}$. The results are obtained by averaging over 10 different trajectories, each contains $10^6$ data points. All the LD simulations are performed using LAMMPS package~\!\cite{lammps}. 

It is worth stressing that all results presented here hold (qualitatively) by using and over-damped dynamics, instead of Eq.~(\ref{langevin_equ}): 
the diffusion coefficient $D_{\textrm{CM}}$ with respect to the SP force $F_{\textrm{sp}}$ exhibits a non-monotonic behavior, the turning number $\psi$ follows the same dynamics, etc.



\begin{figure*}[t]
\begin{center}
\includegraphics[width=0.9 \textwidth]{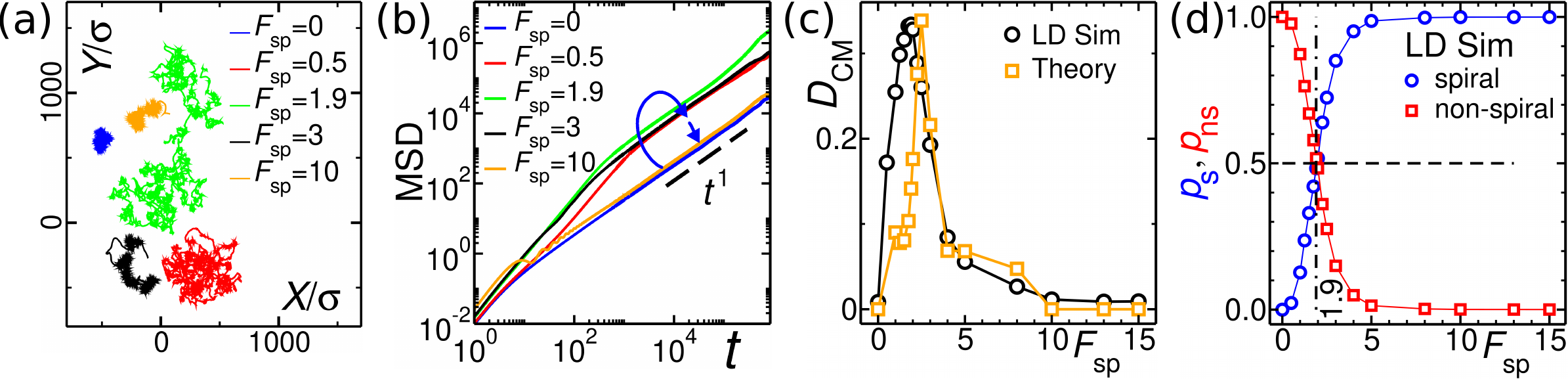}
\end{center}
\caption{
(a) Some typical trajectories of the polymer CM for various values of the SP force $F_{\textrm{sp}}=0$ (blue line), 0.5 (red line), 1.9 (green line), 3 (black line) and 10 (orange line). 
(b) The MSD of the polymer CM as a function of time $t$ for the same values of $F_{\textrm{sp}}$ as of panel (a). The black dashed line $t^1$ is a guide to the eye. The non-monotonic behavior in the MSD is denoted by the blue curved arrow.
(c) Non-monotonic diffusion coefficient of the polymer center of mass $D_{\textrm{CM}}$ as a function of the SP force $F_{\textrm{sp}}$, 
from LD simulations (LD Sim; black circles) and the theory obtained from Eq.~\!(\ref{D_theory}) (Theory; orange squares).
(d) The probability of finding the polymer in the {\it non-spiral} state $p_{\textrm{ns}}$ (red squares) and in the {\it spiral} state $p_{\textrm{s}}$ (blue circles) as a function of the SP force $F_{\textrm{sp}}$. The intersection of the red and blue curves (denoted by the horizontal black dashed line at $p_{\textrm{ns}} = p_{\textrm{s}} = 0.5$ and the vertical black dashed-dotted line at $F_{\textrm{sp}} = 1.9$) corresponds to the peak of the black curve $D_{\textrm{CM}}$ in panel~\!(c). } \label{fig3}
\end{figure*}


\section{Spiral states \& transport}

The self-propelled polymer possesses different configurations that correspond to {\it spiral} and {\it non-spiral} states, which are quantified by considering the turning number $\psi$ of the polymer. The turning number of the entire polymer with $N-1$ bonds is defined as ~\!\cite{turn_handbook}
\begin{equation}
\psi=\frac{1}{2\pi}\sum^{N-1}_{i=1} ( \theta_{i+1}-\theta_{i} ) ,
\label{psi}
\end{equation}
where $\theta_{i}$ is the angle between the $i^{\textrm{th}}$ bond vector and horizontal axis.
In Fig.~\!\ref{fig2}(a) the time evolution of the turning number $\psi (t) $ has been plotted. 
The top left, top right and the bottom insets of  Fig.~\!\ref{fig2} are typical snapshots of the polymer in the {\it non-spiral} state, {\it spiral} state with $\psi>0$ (CCW) and $\psi<0$ (CW), respectively. As mentioned before, in each segment the SP force $F_{\textrm{sp}}$ acts on the green circle directing from the red circle (tail) to the green one (head).
Note that $\psi > 0$ [top right snapshot in Fig.~\!\ref{fig2}(b)] and $\psi < 0 $ [bottom left snapshot in Fig.~\!\ref{fig2}(a)] imply that the global configuration of the polymer is counter-clockwise (CCW) and clockwise (CW), respectively. A vanishing $\psi$ value indicates either a rod-like configuration or a random coil polymer configuration [top left snapshot in Fig.~\!\ref{fig2}(a)].  

In simulations, we find that the probability distribution function (PDF) of the turning number $P(\psi)$ has three peaks: two are located in the left and right hand sides and one at $\psi =0$; see Fig.~\!\ref{fig2}(b) where two vertical black dashed lines, at the two minima of the $P(\psi)$, indicate the region of $\psi$ that corresponds to {\it non-spiral} states. 
The dependency of $P(\psi)$ with the active force $F_{\textrm{sp}}$ is shown in Appendix~\!B. The different types of transition between the {\it non-spiral} and the {\it spiral} states are displayed and discussed in Appendix~\!C.

In the {\it spiral} state, the sum of all active forces acting on different segments of the polymer is almost negligible and thus the center of mass of the polymer barely moves compared to the {\it non-spiral} state. 
On the other hand, when the active polymer escapes from a {\it spiral} configuration (and stays in the {\it non-spiral} state), it is less coiled and active forces along the polymer contribute significantly more to the displacement of the center of mass.  
Fig.~\!\ref{fig3}(a) illustrates some typical trajectories of the polymer CM for different values of the SP force $F_{\textrm{sp}} = 0$ (in blue color), 0.5 (in red color), 1.9 (in green color), 3 (in black color) and 10 (in orange color). Note that the the covered area by the polymer center of mass in a given time first increases and then decreases with the SP force. This is confirmed by the non-monotonic behavior in the MSD as a function of  time; see Fig.~\!\ref{fig3}(b). 
The superdiffusive regime due to the polymer activity at the intermediate time scales is followed by the diffusive regime at long time scales. For large values of the $F_{\textrm{sp}}$ an oscillatory behavior in the MSD at the intermediate time scales is seen (orange line for $F_{\textrm{sp}} = 10$) at which the polymer spends most of its time in the {\it spiral} state.

Fig.~\!\ref{fig3}(c) shows the $D_{\textrm{CM}}$  as a function of $F_{\textrm{sp}}$, computed from the mean squared displacement (MSD) curves in panel~\!(b), for a fixed value of $\ell_{\textrm{p}} = 5$ (black circles), where it is evident that $D_{\textrm{CM}}$  is non-monotonic with respect to the $F_{\textrm{sp}}$. 
We find that there exists an optimal value of $F_{\textrm{sp}}$ that maximizes $D_{\textrm{CM}}$. 
The maximum of $D_{\textrm{CM}}$ takes place when the probabilities of finding the polymer in the {\it spiral} state $P_{\textrm{s}}$ and {\it non-spiral} state $P_{\textrm{ns}}$ are equal to each other; see Fig.~\!\ref{fig3}(d). 
%
\begin{figure*}[]
\centering
\includegraphics[width=0.99 \textwidth]{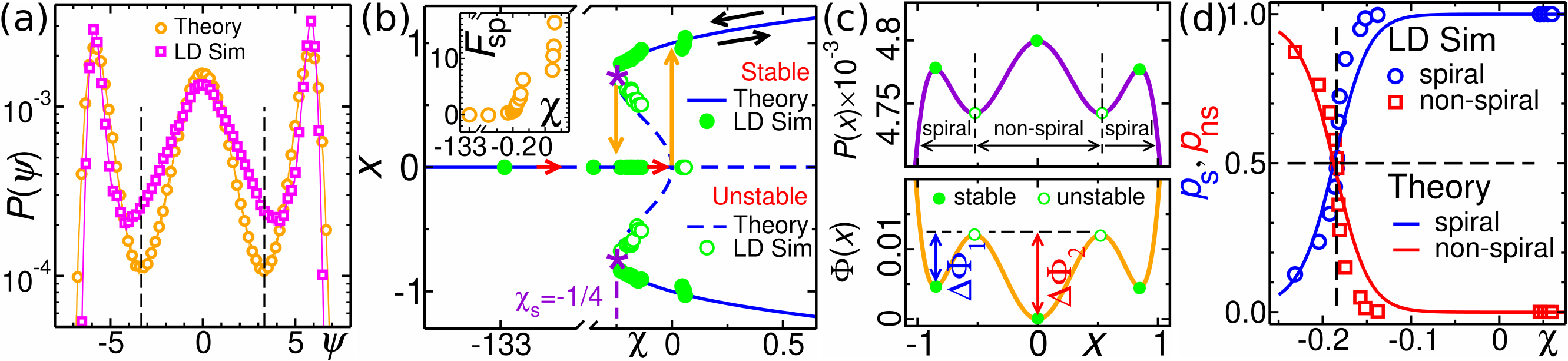}
\caption{(a) The PDF of the $\psi$, $P(\psi)$ obtained from integration of Eq.~\!(\ref{force_non_dimension_theory}) with $\chi=-0.184$ corresponding to SP force $F_{\textrm{sp}}=1.9$ and for fixed value of the noise $\Gamma=0.00375$ (orange circles) and from the LD simulations (purple squared). (b) Bifurcation diagram: theory vs. simulation. The blue solid and dashed lines that are obtained from Eq.~\!(\ref{fixed_points_theory}) show the stable and unstable fixed points, respectively. The filled and empty green circles represent the stable and unstable fixed points, respectively, and are coming from the LD simulations. 
Violet stars denote two additional saddle-node bifurcations at $\chi_{\textrm{s}}=-1/4$.
Inset presents the value of $F_{\textrm{sp}}$ as a function of $\chi$. (c) The PDF of $x$, $P(x)$ (top) and the corresponding potential $\Phi (x)$ (bottom). In both panels the value of $\chi$ has been set to $\chi=-0.2$. (d) The probability of finding the polymer in the {\it non-spiral} state $p_{\textrm{ns}}$ (red squares from the LD simulation and red solid line from the theory) and in the {\it spiral} state $p_{\textrm{s}}$ (blue circles from the LD simulation and blue solid line from the theory) as a function of $\chi$. The intersection of the red and blue curves (denoted by the horizontal black dashed line at $p_{\textrm{ns}} = p_{\textrm{s}} = 0.5$ and the vertical black dashed-dotted line at $\chi = -0.184$ corresponds to $F_{\textrm{sp}}=1.9$) corresponds to the peak of the black curve $D_{\textrm{CM}}$ in Fig.~\!\ref{fig3}(c).}
\label{fig4}
\end{figure*}
These probabilities are defined as $p_{\textrm{s}} = \frac{\langle \tau_{\textrm{s}} \rangle} { \langle \tau_{\textrm{s}} \rangle +  \langle \tau_{\textrm{ns}}  \rangle}$ and $p_{\textrm{ns}} = \frac{\langle \tau_{\textrm{ns}} \rangle} { \langle \tau_{\textrm{s}} \rangle +  \langle \tau_{\textrm{ns}}  \rangle}$, where $\langle \tau_{\textrm{s}} \rangle$ and $\langle \tau_{\textrm{ns}} \rangle$ refer to the average time the polymer stays in the {\it spiral} and {\it non-spiral} states, respectively. 


\section{Theory}
\label{theory}

To analyze the structural dynamics of the polymer, we observe that in the steady state the PDF of the turning number $\psi$, $P(\psi)$, can be empirically described by $P(\psi)\sim \textrm{exp}[-\Phi(\psi)]$ with $\Phi(\psi) = a_{2}\psi^{2} - a_{4}\psi^{4} + a_{6}\psi^{6}$  a potential, where $a_{2}, a_{4}$ and $a_{6}$ are constants that depend on the value of $F_{\textrm{sp}}$ and are obtained using the best fitting of the PDF of the turning number coming from LD simulation by the $P(\psi)$ [see Fig.~\ref{fig2}(b), and also Fig.~\!\ref{fig6} in Appendix~B]; cf. Ref.~\!\citenum{turning}. 
Let us recall that given a Langevin equation $\dot{x}=-\frac{\partial u}{\partial x}+\xi$, with $\xi$ a standard white noise (with intensity $1$), 
the steady state solution of the corresponding Fokker-Planck equation is $P(x)\sim  \textrm{exp}[-u(x)]$. 
We use this results to expressed the dynamics of $\psi$, in dimensionless units, as:
\begin{equation}
\frac{d x}{d \tau} = f(x) + \xi ({\tau})= \chi x + x^{3} - x^{5}+ \xi ({\tau}),
 \label{force_non_dimension_theory}
\end{equation}
where $f(x)  = - d \Phi(x) / d x$ is an effective ``force", 
$x=\psi/U$ and $\tau = t/\mathcal{T}$ with $U^{2}=2a_{4}/(3a_{6})$, $\mathcal{T} = 3a_{6}/(8a^{2}_{4})$ and $\chi = -3a_{2}a_{6}/(4a^{2}_{4})$. 
The values of $a_2$, $a_4$ and $a_6$ as a function of $F_{\textrm{sp}}$ and the corresponding values of $\chi$, $\mathcal{T}$ and $U$ can be found in Table~\!\ref{tab:table1}.
%
\begin{table}[t]
  \begin{center}
    \caption{Fitting parameters for $\ell_{p}=5$.}
    \label{tab:table1}
    \resizebox{8cm}{!}{
    \begin{tabular}{|c|c|c|c|c|c|c|}
       \hline
       $F_{\textrm{sp}}$ & $a_2$ & $a_4$ & $a_6$ & $\chi$ & $\mathcal{T}$ & $U$\\
       
      \hline \hline		
	   0.0  & 1.337 & 0.00348 & 0.00161   & -132.947 & 49.715 & 1.200 \\ \hline
	   0.5  & 0.372 & 0.0205  & 0.000534  & -0.353 &  0.473 & 5.065 \\ \hline
	   1.0  & 0.307 & 0.0164  & 0.0002729 & -0.231 & 0.376 & 6.344 \\ \hline
	   1.25 & 0.362 & 0.0213  & 0.000342  & -0.204 & 0.282 & 6.443 \\ \hline
	   1.5  & 0.447 & 0.0272  & 0.000424  & -0.192 & 0.214 & 6.539 \\ \hline
	   1.75 & 0.569 & 0.0346  & 0.000523  & -0.186 & 0.163 & 6.641 \\ \hline
	   1.9  & 0.736 & 0.0444  & 0.000658  & -0.184 & 0.125 & 6.706 \\ \hline
	   2.0  & 0.857 & 0.0512  & 0.000749  & -0.183 & 0.107 & 6.751 \\ \hline
	   2.25 & 1.153 & 0.0677  & 0.000966  & -0.182 & 0.0789 & 6.837 \\ \hline
	   2.5  & 1.233 & 0.0716  & 0.00100   & -0.180 & 0.0733 & 6.900 \\ \hline
	   3.0  & 0.823 & 0.0492  & 0.000684  & -0.174 & 0.105 & 6.927 \\ \hline
	   4.0  & 0.497 & 0.0336  & 0.000476  & -0.1573 & 0.158 & 6.856 \\ \hline
	   5.0  & 0.490 & 0.0351  & 0.000510  & -0.151 & 0.154 & 6.776 \\ \hline
	   8.0   & 0.45  & 0.0350  & 0.00050    & -0.137755  & 0.153 & 6.831 \\ \hline
	   10.0 & -0.150  & 0.0350  & 0.00050   & 0.0459 & 0.1531 & 6.831 \\ \hline
   	   13.0  & -0.165 & 0.0350  & 0.00050  & 0.05051 & 0.15306 & 6.831 \\ \hline
	   15.0 & -0.180  & 0.0350  & 0.00050   & 0.0551 & 0.1531 & 6.831 \\ \hline
	   20.0  & -0.195 & 0.0350  & 0.00050  & 0.0597 & 0.1531	& 6.831 \\ \hline
      
    \end{tabular}
    }
  \end{center}
\end{table}
%
The term $\xi$ refers to a white noise that satisfies $\langle \xi(\tau)\rangle = 0$ and $\langle \xi(\tau) \xi(\tau^{'})\rangle =2\Gamma\delta(\tau - \tau^{'})$.
The noise intensity $\Gamma$ is obtained by mapping the PDF of the turning number in LD simulations with the one resulting from the integration of Eq.~\!(\ref{force_non_dimension_theory}) [see Fig.~\!\ref{fig4}(a)] and has a  fixed value of $\Gamma = 0.00375$ for different values of $F_{\textrm{sp}}$. 
Note that though the noise in Eq.~(\ref{force_non_dimension_theory}) was assumed white, a colored noise is, arguably, a more generic option. 
In short, the dynamics of $\psi$ can be expressed, instead of using Eq.~(\ref{force_non_dimension_theory}), 
by a  generalized Langevin equation with the memory kernel\cite{Panja_J_Stat_Mech,Vandebroek_J_Stat_Phys} and/or a mobility kernel~\cite{Sakaue_PRE_2013,Saito_PRE_2015}, or a combination of both~\cite{Shinkai}. Let us recall that memory effects occur, for instance, in protein folding~\cite{Polymer_folding_PRE_2102,Polymer_folding_PRL_2104,Soft_Matter_2017}  and in  polymer translocation~\cite{Poly_trans_J_Cond_Matt_2013,Poly_trans_Soft_Matter_2014,Poly_trans_Polymers_2016}, where tension propagation theory provides a reliable description~\cite{Sakaue_PRE_2007,Rowghanian_2011,Timo_PRE_2012,Jalal_JCP_2014,Jalal_JCP_2015,Jalal_EPL_2017, Jalal_Sci_Rep_2017,Jalal_Cond_Matt_2018,Jalal_Cond_Matt_2020,Jalal_PRR_2021,Jalal_PRR_2023}. 
However, for  the dynamics of $\psi$, as we show below, it is sufficient with Eq.~(\ref{force_non_dimension_theory}) and a white noise:  Eq.~(\ref{force_non_dimension_theory}) does not only allow us to obtain the correct steady state of $P(\psi)$, it also captures 
the correct transition rates between spiral and non spiral configurations.

The deterministic form of Eq.~\!(\ref{force_non_dimension_theory}) -- that corresponds to $\Gamma=0$ and thus $\dot{x} = f(x;\chi)$ --  undergoes a {\it subcritical pitchfork} bifurcation~\!\cite{strogatz}; the time-dependent solution of  Eq.~\!(\ref{force_non_dimension_theory}) is shown in Appendix~D. 
The bifurcation diagram in the $x^{*}-\chi$ plane is shown in Fig.~\!\ref{fig4}(b).  
The fixed points of Eq.~\!(\ref{force_non_dimension_theory}) -- $f(x^{*};\chi)=0$ -- are: 
\begin{equation}
x^{*} = 0 ,\, \pm \bigg( \frac{1 \pm \sqrt{1+4\chi}}{2} \bigg)^{1/2}\, .
\label{fixed_points_theory}
\end{equation}
For $\chi>0$, $x^{*}=0$ is an (linearly) unstable fixed point since  $f^{'}(x^{*})=\frac{\partial f}{\partial x}(x^{*})\geq0$. 
For $\chi\leq0$, $x^{*}=0$ becomes stable ($f^{'}(0)<0$) and two unstable fixed points -- such that $f^{'}(x^{*})\geq0$ -- emerge.  
At $\chi_{\textrm{s}}=-1/4$ there are two additional saddle-node bifurcations [denoted by violet stars in the main panel of Fig.~\!\ref{fig4}(b)] which are obtained by considering the condition $ 1 + 4 \chi_{\textrm{s}} \geq 0$.  
Note that the fixed points are the  extremum values of the $P ( \psi )$ [see Fig.~\!\ref{fig6} in the Appendix~B]. 
The effect of $x^{5}$ term in Eq.~\!(\ref{force_non_dimension_theory}) in the bifurcation diagram is to turn around the unstable branches at $\chi = \chi_{\textrm{s}}$ and become stable for $\chi > \chi_{\textrm{s}}$ [solid blue lines in the top and bottom in Fig.~\!\ref{fig4}(b) main panel]. These stable states provide the possibility of {\it jumps} and {\it hysteresis} as $\chi$ is varied. The filled and empty green circles in Fig.~\!\ref{fig4}(b) correspond to the LD simulations and show the stable and unstable states, respectively. The simulations data and theoretical curves show a perfect agreement. 
The inset in Fig.~\!\ref{fig4}(b) shows how the values of $F_{\textrm{sp}}$ and $\chi$ relate to each other.

To compute the diffusion coefficient $D_{\textrm{CM}}$, and to put in evidence the {\it run-\&-tumble} character of the process, we consider that during the {\it spiral} states, the sum of all active forces acting on different segments of the polymer is negligible. Thus, in this state, the speed (of the CM) of polymer is approximately $v_{\textrm{s}}\simeq 0$. 
This implies that we assume that the distance travelled by the polymer in the {\it spiral} sates is negligible [see the movie in SM]. 
On the other hand, in the {\it non-spiral} states, the polymer is (in comparison to {\it spiral} states) extended, 
and thus, the sum of active forces is non-zero. 
In consequence, in the {\it non-spiral} state, the polymer moves at  speed $v_{\textrm{ns}}>0$, 
and we assume that it travels an average distance $\ell=v_{\textrm{ns}} \langle \tau_{\textrm{ns}} \rangle$. 
To approximately determine $v_{\textrm{ns}}$, let us assume that  in the {\it non-spiral} state, the polymer is an extended rod -- that experiences an active force $NF_{\textrm{sp}}/\ell_\textrm{p}$ and a friction drag $N \, \gamma$  -- and thus, moves at speed $v_{\textrm{ns}}=F_{\textrm{sp}}/\gamma  ~\! \ell_\textrm{p}$. 
All together, this implies that as the polymer goes through a cycle {\it spiral}-{\it non-spiral}, i.e. involving a transition from the {\it spiral} to the 
{\it non-spiral} state, moves an average distance $\ell$. Since the average duration of these cycles is 
$\Omega=\langle \tau_{\textrm{s}} \rangle+\langle \tau_{\textrm{ns}} \rangle$, in a time $t$, the polymer performs an average of $n=t/\Omega$ of those cycles. 
%
%
It is worth noting that  $\langle \tau_{\textrm{ns}}\rangle$ and $\langle \tau_{\textrm{s}}\rangle$ can be estimated from the transition rates to escape from the {\it non-spiral} and {\it spiral} states considering the associated Kramer's escape problem that assumes these rates are proportional to the exponential of the potential barrier:
$\langle \tau_{\textrm{ns}} \rangle^{-1} = A_{\textrm{ns}} \exp(\Delta \Phi_1/\Gamma)$ and 
$\langle \tau_{\textrm{s}} \rangle^{-1} = A_{\textrm{s}} \exp(\Delta \Phi_2/\Gamma)$, respectively~\!\cite{van_Kampen}, where $\Delta \Phi_1= \frac{1}{12}(1+4\chi)^{3/2}$ 
and $\Delta \Phi_2= \frac{1}{48}(-1+\sqrt{1+4\chi})(1+8\chi-\sqrt{1+4\chi})$ [see Fig.~\!\ref{fig4}(c) bottom panel]. 
The detail of how $\Delta \Phi_1$ and $\Delta \Phi_2$ are obtained, can be seen in the Appendix~D. 
The estimates for $A_{\textrm{ns}}$ and $A_{\textrm{ns}}$ in Fig.~\!\ref{fig3}(c) are $A_{\textrm{ns}}=A_{\textrm{s}}=2.35$.
Recall that then $p_{\textrm{s}}$ and $p_{\textrm{ns}}$ can also be expressed in term of these rates: $p_{\textrm{s}} = \langle \tau_{\textrm{s}} \rangle/\Omega$ and  $p_{\textrm{ns}} = \langle \tau_{\textrm{ns}} \rangle/\Omega$.  
This is shown in  Fig.~\!\ref{fig4}(d). Note that there is a good agreement between estimates from LD simulations and these theoretical estimates.

Finally, let us recall that for a random walk (RW) after $n_{\textrm{RW}}$ steps of length $\ell_{\textrm{RW}}$, the MSD is written as $\langle \mathbf{r}^2 \rangle = n_{\textrm{RW}}\,\ell_{\textrm{RW}}^{2}$~\!\cite{Reif}. 
Thus, if we interpret that in time $t$, the $n$ cycles performed by the active polymer correspond to 
$n_{\textrm{RW}}$ steps of a RW, that moves in each step an average distance $\ell_{\textrm{RW}}=\ell$, we
can approximate the diffusion coefficient -- using the above-given expressions for $n$ and $\ell$ -- as: 
\begin{equation}
D_{\textrm{CM}} = \lim_{t\to\infty} \frac{\langle \mathbf{r}^2 \rangle}{4 t} = \frac{v_{\textrm{ns}}^2}{4} \frac{\langle \tau_{\textrm{ns}} \rangle^2}{\langle \tau_{\textrm{s}} \rangle+\langle \tau_{\textrm{ns}} \rangle} = \frac{\overline{v}^2\Omega}
{4} \, ,
\label{D_theory}
\end{equation}
where $\overline{v}=p_{\textrm{ns}}v_{\textrm{ns}}$, and is in agreement with the run-and-tumble particle model~\!\cite{Sokolov_PRE_2012}.
The comparison between the diffusion coefficients $D_{\textrm{CM}}$ predicted by Eq.~\!(\ref{D_theory}) (denoted by "Theory"; orange squares) and the one obtained from the LD simulations ("LD Sim"; black circles) displayed in Fig.~\!\ref{fig3}(c), shows that there exists a  good qualitative agreement between LD simulations and the theory. 

Importantly, Eqs.~\!(\ref{force_non_dimension_theory}) and (\ref{D_theory}) allow us to prove that there is an optimal self-propelling force that maximizes the diffusion coefficient. 
The existence of such an optimal force value is easy to understand. When $F_{\textrm{sp}}\to 0$, the spiral states is not observed. The speed in the non-spiral state is $v_0 \propto F_{\textrm{sp}}$, then, $F_{\textrm{sp}} \to 0$ implies $D_{\textrm{CM}} \to 0$. 
On the contrary, for $F_{\textrm{sp}} \gg 1$, the polymer is locked in spirals, but since we assume that $v_0 ~\! \simeq ~ 0$ in this state, we get again \ $D_{\textrm{CM}} \to 0$. 
In summary, increasing $F_{\textrm{sp}}$  increases $v_0$, but at the same time enhances the probability of finding the polymer in the spiral state, where the motion of the center of mass of the polymer is strongly reduced. 
And thus, we find that for intermediate values of $F_{\textrm{sp}}$, $D_{\textrm{CM}}$ is maximized. Compare the results obtained here with those reported in a numerical study with the Krotky-Porod model~\!\cite{theo4}.

\section{Conclusions}

Our study puts in evidence the existence of a key coupling between the structural dynamics of the polymer and its transport properties. 
Moreover, we have shown that there exists an optimal active force $F_{\textrm{sp}}$ that maximizes the diffusion coefficient 
$D_{\textrm{CM}}$. 
It would be interesting to employ the optimal Langevin modeling of out-of-equilibrium molecular dynamic simulations and compare it with the approach developed here~\!\cite{OptimalLangevin}.
The reported results are of relevance for a large number 
of experimental active polymer studies, including actin filament motion in motility assays experiments, among many other examples.   

\begin{figure}
\includegraphics[width=0.9 \columnwidth]{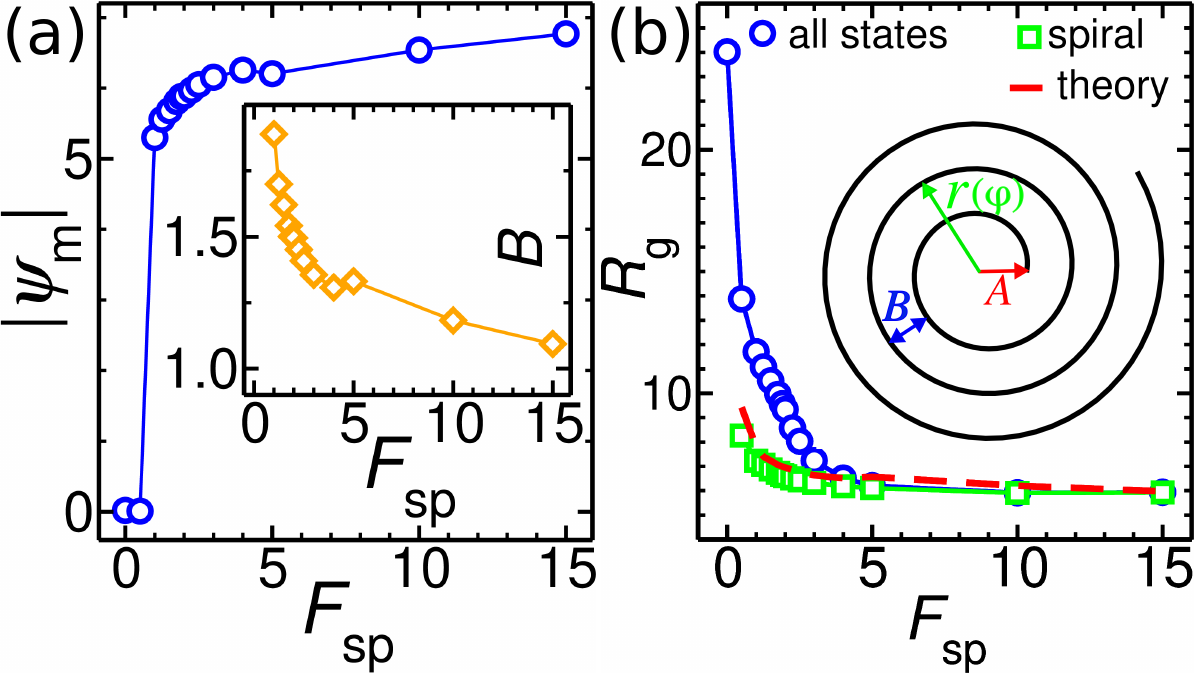}
\caption{(a) The absolute value of the maximum turning number $|\psi_{m}|$ as a function of $F_{\textrm{sp}}$ (blue circles) corresponding to the left and right peaks in Fig.~\!\ref{fig6}. Inset shows the value of $B$ obtained from the Eq.~\!(\ref{turning_number_theory}) as a function of $F_{\textrm{sp}}$ by employing the values of the $|\psi_{m}|$ from the main panel. 
(b) The radius of gyration of the polymer $R_{\textrm{g}}$ as a function of $F_{\textrm{sp}}$ for fixed value of the persistence length $\ell_{p}=5$, and for mixture of both {\it non-spiral} and {\it spiral} states (denoted by all states, blue circles), for just {\it spiral} state (green squares), and obtained from Eq.~\!(\ref{rg2_continous_extended}) by the theory (red dashed line). Inset shows the {\it Archimedean} spiral.}
\label{fig5}
\end{figure}

\section{Appendix A: Radius of gyration and turning number; theory \& simulation}
\label{Gyration_Rad}

The spatial size of the polymer in the {\it spiral} state, specifically radius of gyration (RG), is related to $\psi$. 
Assume that the polymer can be approximately described by an {\it Archimedean} spiral [inset of Fig.~\!\ref{fig5}(b)] in which the radius $r$ is a function of the angle $\varphi$ such that  $r(\varphi)=A+B\varphi / (2 \pi)$. 
The coefficients $A$ and $B$ are functions of $F_{\textrm{sp}}$ and $\ell_\textrm{{p}}$.
Under this assumption, the RG of polymer $R_{\textrm{g}}$ with contour length of $N$ is computed as ~\!\cite{rubin_stein}
\begin{equation}
R_{\textrm{g}} = \{ \frac{1}{N}\int_{0}^{N} \left[\vec{r}(u) - \vec{r}_{\textrm{CM}} \right]^{2} du \}^{1/2},
\label{rg_integral}
\end{equation}
where $\vec{r}(u)$ and $\vec{r}_{\textrm{CM}}$ are the position vector of the monomers and the position vector of the CM, respectively.
By setting $\vec{r}_{\textrm{CM}} = \vec{0}$ as the origin and using the change in the integral variable $d u = \lambda d s$, with $\lambda = N / \int d s$ as the linear monomer density that is set to unity, the RG is written as 
\begin{equation}
R_{\textrm{g}} = \bigg\{  \frac{\pi}{2 B N} \big[   \big( A + B \psi_{\textrm{m}} \big)^4  - A^4 \big] \bigg\}^{1/2} \,  ,
\label{rg2_continous_extended}
\end{equation}
where $\psi_{\textrm{m}}$ corresponds to the maximum of the turning number, which is a function of $F_{\textrm{sp}}$. 
To obtain $\psi_{\textrm{m}}$, we use the following procedure. Integrating over the length element on the spiral in polar coordinates $ds=r ~ d\varphi$ gives the contour length of the polymer $N$ as a function of $\psi_{\textrm{m}}$ as $\int_0^{N} ds = \int_{0}^{2\pi \psi_{\textrm{m}}} r(\varphi) d\varphi$. Here, all quantities have been written in the LJ unites. 
Finally, by integrating both sides of the above equality, we arrive at

\begin{equation}
\pi B \psi_{\textrm{m}}^{2} + 2\pi A \psi_{\textrm{m}} - N = 0.
\label{turning_number_theory}
\end{equation}

Then, $\psi_{\textrm{m}}$ is obtained from the roots of Eq.~\!(\ref{turning_number_theory}). In the main panel (a) in Fig.~\!\ref{fig5} the absolute value of the maximum turning number $| \psi_{\textrm{m}} | $ has been plotted as a function of $F_{\textrm{sp}}$, that are obtained from the LD simulations by considering the most probable values of $\psi$ (left and right peaks in Fig.~\!\ref{fig6}). The inset shows the value of the coefficient $B$ as a function of $F_{\textrm{sp}}$. In Fig.~\ref{fig5}(b) using the LD simulations the $R_{\textrm{g}}$ has been plotted as a function of $F_{\textrm{sp}}$ when contributions of both {\it non-spiral} and {\it spiral} states have been taken into account (blue circles) and also when only the {\it spiral} state has been taken into account (green diamonds). As seen, the RG is a monotonic and decreasing function of the SP force due to the increasing of probability of finding the polymer in the {\it spiral} state [see Fig.~\!\ref{fig3}(d)]. Substituting the values of $A=1$ and $B$ [presented in the inset of Fig.~\!\ref{fig5}(a)] into Eq.~\!(\ref{rg2_continous_extended}), the semi-analytical values of the RG as a function of SP force is plotted in Fig.~\!\ref{fig5}(b) (red dashed line). As seen the semi-analytical data are in a very good agreement with those coming from the LD simulations for the {\it spiral} state.\\


\section{Appendix B: PDF of turning number}
\label{Turning_num}

\begin{figure}[bt]
\begin{center}
\includegraphics[width=0.5\columnwidth]{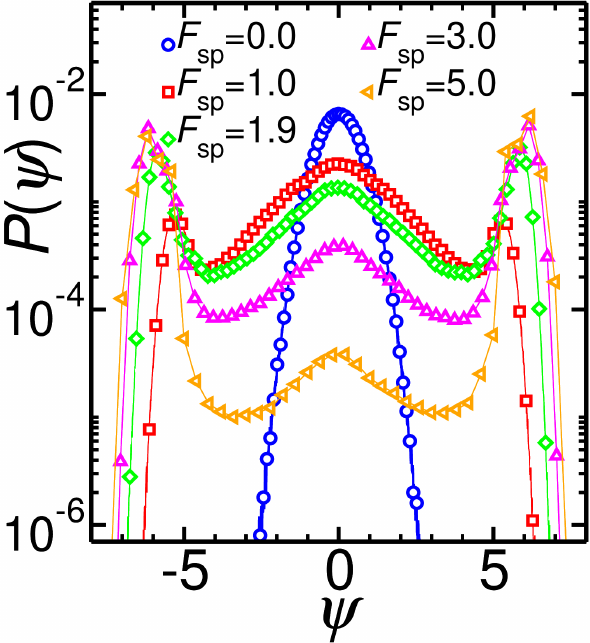}
\end{center}
\caption{
The PDF of the turning number $P ( \psi )$ with fixed values of the polymer contour length of $N=200$, persistence length of $\ell_{\textrm{p}} = 5 $, for various values of of the SP forces $F_{\textrm{sp}}=$0.0 (blue circles), 1.0 (red squares), 1.9 (green diamonds), 3.0 (pink triangles-up) and 5.0 (orange triangles left).
} 
\label{fig6}
\end{figure}

Fig.~\!\ref{fig6} represents the PDF of the turning number $P ( \psi )$ for an active polymer with contour length of $N=200$, persistence length of $\ell_{\textrm{p}} = 5 $, for various values of the SP forces $F_{\textrm{sp}}=$0.0 (blue circles), 1.0 (red squares), 1.9 (green diamonds), 3.0 (pink triangles-up) and 5.0 (orange triangles left).

\section{Appendix C: Transitions between the {\it non-spiral} and the {\it spiral} states}
\label{Transitions}

\begin{figure}
\includegraphics[width=0.8 \columnwidth]{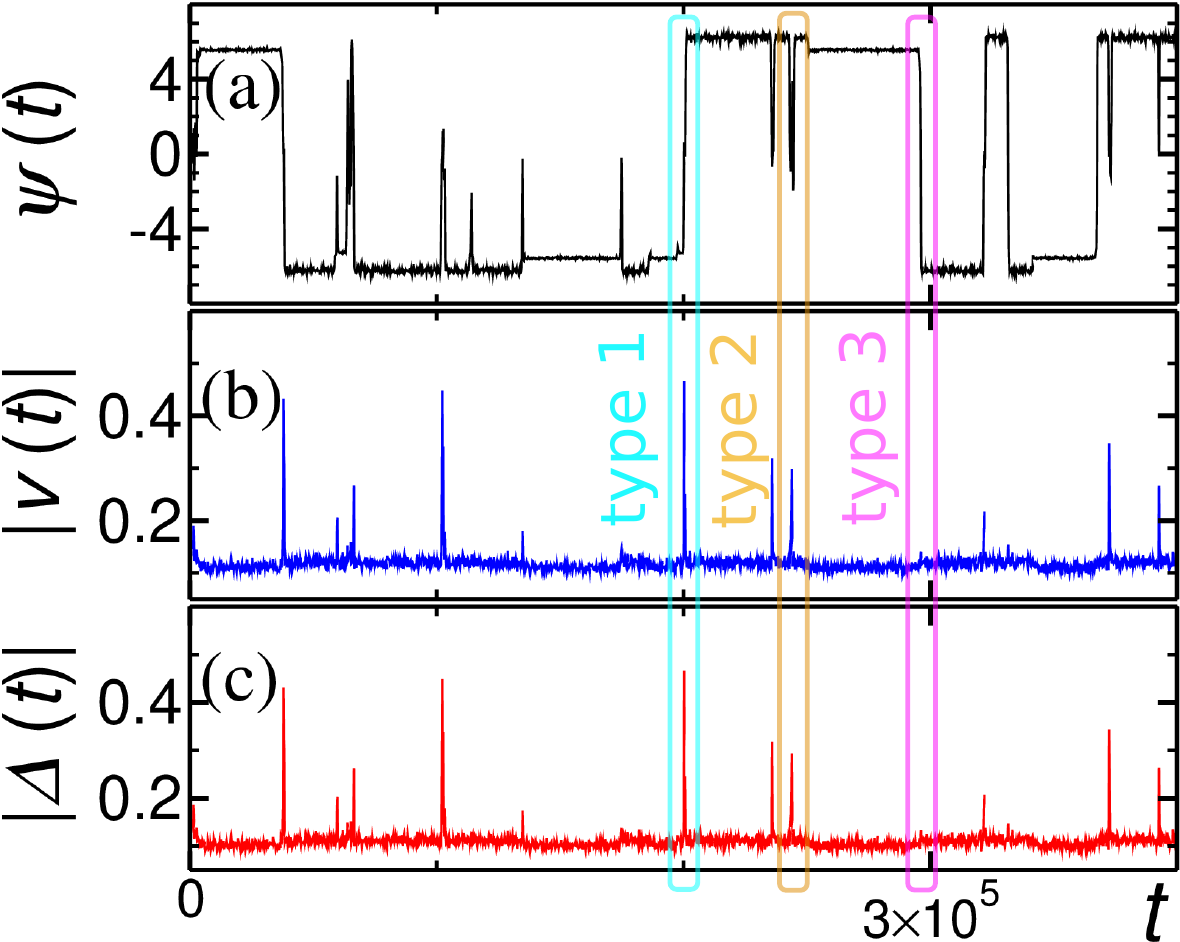}
\caption{Three types of transition between the {\it non-spiral} and the {\it spiral} states. (a) Turning number as a function of time. Panels (b) and (c) are the speed and displacement of the polymer CM, respectively, as a function of time corresponding to panel (a). 
Each type of transition is distinguished by a different color box (see the text in the Appendix~C for more details). The values of the system parameters are $N=200$, $\ell_{\textrm{p}} = 5 $ and $F_{\textrm{sp}}=4$.} 
\label{fig7}
\end{figure}

In general, there are three different types of transitions between the {\it non-spiral} and the {\it spiral} states. In panel (a) of Fig.~\!\ref{fig7} the time evolution of the turning number $\psi(t)$ has been plotted. Panels (b) and (c) are devoted to present the corresponding speed of the polymer CM $\vert v(t) \vert$ and displacement of the polymer CM $\vert \Delta(t) \vert$, respectively. Three different types of the transitions are distinguished by the three vertical color boxes. 
The type $1$ (cyan box) is a {\it spiral-non-spiral-spiral} transition, in which the polymer transients from a CW {\it spiral} state to a {\it non-spiral} state, and then to a CCW {\it spiral} state. The inverse transition is also possible, i.e. from a CCW {\it spiral} state to a {\it non-spiral} state, and then to a CW {\it spiral} state (not shown here). 
In type $2$ (orange box), the polymer is opened up from a CCW {\it spiral} state to a {\it non-spiral} state, and then gets back to a CCW {\it spiral} state. Obviously, the transition CW-{\it non-spiral}-CW is also possible. 
In type $3$ (magenta box), the polymer directly transients from a CCW {\it spiral} state to a CW {\it spiral} state without passing from an intermediate {\it non-spiral} state, again with possibility of the inverse transition. 
In type $3$, the polymer starts to be opened up from its interior layers and changes its direction of rotation without any passing through an intermediate {\it non-spiral} transient state. 
In panels (b) and (c) of Fig.~\!\ref{fig7}, the peaks represent the transition to the {\it non-spiral} state. As seen, there is not any peak for the type $3$ transition due to the direct transition between CCW and CW {\it spiral} states without any passing through an intermediate {\it non-spiral} transient state. To show the three types of the transition a typical movie has been provided in the SM.

\begin{figure}[t]
\centering
\includegraphics[width=0.99 \columnwidth]{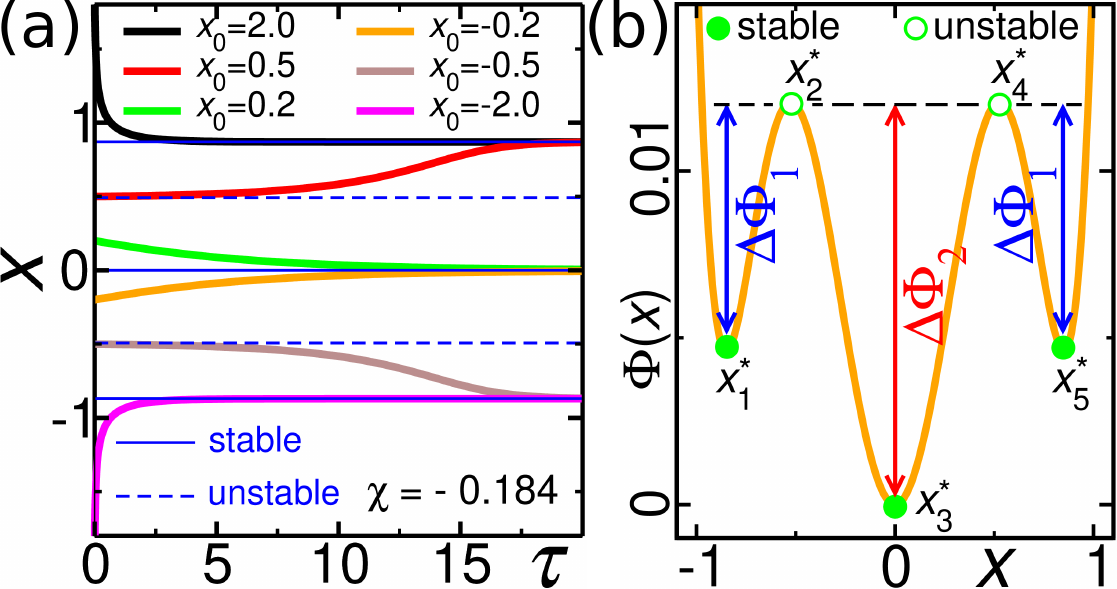}
\caption{(a) Time evolution of the solution (the normalized turning number $x$) of Eq.~\!(\ref{time_evolution}) for different values of the initial normalized turning number $x_{0}$ and for fixed value of $\chi=-0.184$ corresponding to $F_{\textrm{sp}}=1.9$. The solid and dashed blue lines represent the stable and unstable fixed points corresponding to the $\chi=-0.184$, respectively. 
(b) The potential $\Phi(x)$ as a function of $x$ for $\chi=-0.2$. The filled and open green circles represent the stable and unstable extrema (fixed points) of the $\Phi(x)$, respectively. The $\Delta\Phi_{1}$ and $\Delta\Phi_{2}$ are the potential barriers that the polymer has to overcome to make transitions between {\it spiral} and {\it non-spiral} states.}
\label{fig8}
\end{figure}


\section{Appendix D: Bifurcation and potential barriers}
\label{Bifurcation}

The deterministic form of the time evolution of the normalized turning number $x$ is described by the following relation (see also the text in Sec.~\!\ref{theory} for more details)
\begin{equation}
\frac{d x}{d \tau} = f(x) = \chi x + x^{3} - x^{5}.
 \label{time_evolution}
\end{equation}
Fig.~\!\ref{fig8}(a) shows that the normalized turning number $x$ asymptotically approaches to a stable fixed point depending on the initial value of $x$. Fig.~\!\ref{fig8}(a) presents the numerical solutions of Eq.~\!(\ref{time_evolution}) for fixed value of $\chi=-0.184$ corresponding to $F_{\textrm{sp}}=1.9$, and for various initial values of the normalized turning number $x_0 = 2.0$ (in black color), $0.5$ (in red color), $0.2$ (in green color), $-0.2$ (in orange color), $-0.5$ (in brown color) and $-2.0$ (in pink color). The blue solid lines represent the stable fixed points at $x^* = 0.87, 0.0$ and $-0.87$, while the blue dashed lines show the unstable fixed points at $x^* = 0.493$ and $-0.493$.\\

Fig.~\!\ref{fig8}(b) shows the potential $\Phi(x)$ as a function of $x$ for fixed value of $\chi=-0.2$. The filled green circles (located at $x^{*}_{1}, x^{*}_{3}$ and $x^{*}_{5}$) and open green circles (located at $x^{*}_{2}$ and $x^{*}_{4}$) represent the stable and unstable fixed points of the system, respectively, that are given by solving $f(x^*)=-\partial \Phi(x)/\partial x|_{x^*}=0$, as $x^*_{1}=-[(1+\sqrt{1+4\chi})/2]^{1/2}, x^*_{2}=-[(1-\sqrt{1+4\chi})/2]^{1/2}, x^*_{3}=0, x^*_{4}=[(1-\sqrt{1+4\chi})/2]^{1/2}$ and $x^*_{5}=[(1+\sqrt{1+4\chi})/2]^{1/2}$. 
To make transitions between the {\it spiral} and {\it non-spiral} states, the polymer has to overcome the potential barriers denoted by the $\Delta \Phi_{1}$ and $\Delta \Phi_{2}$ in Fig.~\!\ref{fig8}(b). It is noticeable that the region between $x^{*}_{2}$ and $x^{*}_{4}$ corresponds to the {\it spiral} state and otherwise to the {\it non-spiral} state (for more detail see the text in Sec.~\!\ref{theory}). As it can be seen in Fig.~\!\ref{fig8}(b), the potential barriers are given by $\Delta\Phi_{1}=|\Phi(x^{*}_{2})-\Phi(x^{*}_{1})|=|\Phi(x^{*}_{4})-\Phi(x^{*}_{5})|$ and $\Delta\Phi_{2}=|\Phi(x^{*}_{2})-\Phi(x^{*}_{3})|=|\Phi(x^{*}_{4})-\Phi(x^{*}_{3})|$,  where $\Phi(x)=-\chi x^{2}/2 - x^{4}/4 + x^{6}/6$. Therefore, the potential barriers are written as $\Delta\Phi_{1}=\frac{1}{12}(1+4\chi)^{3/2}$ and $\Delta\Phi_{2}=\frac{1}{48}(-1+\sqrt{1+4\chi})(1+8\chi-\sqrt{1+4\chi})$.

\section*{Acknowledgements}
FP acknowledges fruitful discussions with Debasish Chaudhury. 
JS acknowledges Iran National Science Fundation "This work is based upon research funded by Iran National Science Fundation (INSF) under project No.~\!4026895".




\begin{thebibliography}{9}


\bibitem{albert}
B. Alberts, J. Alexander, L. Julian, R. Martin, R. Keith, and W. Peter, {\it Molecular Biology of the Cell} (Garland Science, New York, 2002).

\bibitem{zia}
R. K. P. Zia, J. J. Dong, and B. Schmittmann, J. Stat. Phys. \textbf{144}, 405 (2011).

\bibitem{chromatin1}
I. Bronstein, Y. Israel, E. Kepten, S. Mai, Y. Shav-Tal,
E. Barkai, and Y. Garini, Phys. Rev. Lett. \textbf{103}, 018102 (2009).

\bibitem{chromatin2}
I. Bronstein, E. Kepten, I. Kanter, S. Berezin, M. Lind-
ner, A. B. Redwood, S. Mai, S. Gonzalo, R. Foisner,
Y. Shav-Tal, and Y. Garini, Nat . Commun. \textbf{6}, 8044
(2015).

\bibitem{chromatin3}
G. G. Cabal, A. Genovesio, S. Rodriguez-Navarro,
C. Zimmer, O. Gadal, A. Lesne, H. Buc, F. Feuerbach-
Fournier, J.-C. Olivo-Marin, E. C. Hurt,
and U. Nehrbass, Nature \textbf{441}, 770 (2006).

\bibitem{chromatin4}
A. Zidovska, D. A. Weitz, and T. J. Mitchison, Proc.
Natl. Acad. Sci. \textbf{110}, 15555 (2013).

\bibitem{chromatin5}
N. Ganai, S. Sengupta, and G. I. Menon, Nucleic Acids
Res. \textbf{42}, 4145 (2014).

\bibitem{chromatin6}
N. Haddad, D. Jost, and C. Vaillant, Chromosome Research \textbf{25}, 35 (2017).

\bibitem{cytos1}
C. P. Brangwynne, G. H. Koenderink, F. C. MacKintosh,
and D. A. Weitz, Phys. Rev. Lett. \textbf{100}, 118104 (2008).

\bibitem{cytos2}
D. Mizuno, C. Tardin, C. F. Schmidt, and F. C. MacKintosh, Science \textbf{315}, 370 (2007).

\bibitem{turning6}
D. A. Fletcher and R. D. Mullins, Nature \textbf{463}, 485(2010).

\bibitem{turning7}
F. Jülicher, S. W. Grill, and G. Salbreux, Reports Prog.
Phys. \textbf{81}, 076601 (2018).

\bibitem{Motility_Assay_Book}
J. M. Scholey, {\it Motility Assays for Motor Proteins} (Academic Press, New York, 1993).

\bibitem{Chaudhuri_Nano_Lett_2018}
S. Chaudhuri, T. Korten, S. Korten, G. Milani, T. Lana, G. T. Kronnie and S. and Diez, Nano Lett. \textbf{18}, 117 (2018).

\bibitem{Inoue_Nat_Commu_2015}
D. Inoue, T. Nitta, A. Kabir, K. Sada, J. P. Gong, A. Konagaya and A. Kakugo, Nat. Commun. \textbf{7}, 12557 (2016).

\bibitem{Dekker_NanoLett_2005}
M. G. L. van den Heuvel, C. T. Butcher, R. M. M. Smeets, S. Diez and C. Dekker, Nano Lett. \textbf{5},  1117 (2005).

\bibitem{Maximov_NanoLett_2013}
V. Schroeder, T. Korten, H. Linke, S. Diez and I. Maximov, Nano Lett. \textbf{13}, 3434 (2013).

\bibitem{Dekker_Science_2006}
M. G. L. van den Heuvel, M. P. de Graaff and C. Dekker, Science \textbf{312}, 910 (2006).

\bibitem{flagella0}
S. Camalet, F. Jülicher and J. Prost, Phys. Rev. Lett. \textbf{82}, 1590 (1999).

\bibitem{flagella1}
R. Chelakkot, A. Gopinath, L. Mahadevan, and M. F.
Hagan, J. R. Soc., Interface \textbf{11}, 20130884 (2014).

\bibitem{flagella2}
J. Elgeti, R. G. Winkler, and G. Gompper, Rep. Prog.
Phys. \textbf{78}, 056601 (2015).


\bibitem{exp1}
R. Dreyfus, J. Baudry, M. L. Roper, M. Fermigier, H. A.
Stone, J. Bibette, Nature \textbf{437}, 862 (2005).

\bibitem{exp2}
V. Schaller, C. Weber, C. Semmrich, E. Frey, A. R. Bausch, Nature \textbf{467}, 73 (2010).

\bibitem{exp3}
Y. Sumino, K. H. Nagai, Y. Shitaka, D. Tanaka, K. Yoshikawa, H. Chat{\'e} and K. Oiwa, Nature \textbf{483}, 448 (2012).

\bibitem{exp4}
L. J. Hill, N. E. Richey, Y. Sung, P. T. Dirlam, J. J. Griebel,
E. Lavoie-Higgins, I.-B. Shim, N. Pinna, M.-G. Willinger,
W. Vogel, J. J. Benkoski, K. Char, J Pyun, ACS Nano \textbf{8}, 3272 (2014).

\bibitem{exp5}
R. Suzuki and A. R. Bausch, Nat. Commun. \textbf{8}, 41 (2017).


\bibitem{theo1}
T. B. Liverpool, Phys. Rev. E \textbf{67}, 031909 (2003).

\bibitem{theo2}
H. Jiang and Z. Hou, Soft Matter \textbf{10}, 1012 (2014).

\bibitem{theo3}
A. Ghosh and N. Gov, Biophys. J. \textbf{107}, 1065 (2014).

\bibitem{theo4}
R. E. Isele-Holder, J. Elgeti and G. Gompper, Soft Matter, 
\textbf{11}, 7181 (2015).

\bibitem{theo5}
D. Sarkar, S. Thakur, Phys. Rev. E \textbf{93}, 032508 (2016).

\bibitem{theo6}
T. Eisenstecken, G. Gompper and R. G. Winkler, Polymers \textbf{8}, 304 (2016).

\bibitem{theo7}
T. Eisensteckena, G. Gompper and R. G. Winklera, J. Chem. Phys. 
\textbf{146}, 154903 (2017).

\bibitem{theo8}
D. Osmanovi{\'c} and Y. Rabin, Soft Matter \textbf{13}, 963 (2017).

\bibitem{theo9}
S. K. Anand and S. P. Singh, Phys. Rev. E \textbf{98}, 042501 (2018).

\bibitem{PRL}
V. Bianco, E. Locatelli, and P. Malgaretti, Phys. Rev.
Lett. \textbf{121}, 217802 (2018).

\bibitem{theo10}
N. Gupta, A. Chaudhuri and D. Chaudhuri, Phys. Rev. E \textbf{99}, 042405 (2019).

\bibitem{baskaran}
 M. S. E. Peterson, M. F. Hagan and A. Baskaran1. J. Stat. Mech. \textbf{2020}, 
 013216, (2020).

\bibitem{theo11}
R. G. Winkler and G. Gompper, J. Chem. Phys. \textbf{153}, 040901 (2020).

\bibitem{turning}
A. Shee, N. Gupta, A. Chaudhuri and D. Chaudhuri, Soft Matter, \textbf{17}, 2120 (2021).

\bibitem{theo12}
C. A. Philipps, G. Gompper and R. G. Winkler, J. Chem. Phys. 
\textbf{157}, 194904 (2022).

\bibitem{Jabbari_PRE}
M. Fazelzadeh, E. Irani, Z. Mokhtari and S. Jabbari-Farouji, Phys. Rev. E \textbf{108}, 024606 (2023).

\bibitem{Jabbari_JCP}
M. Fazelzadeh, Q. Di, E. Irani, Z. Mokhtari and S. Jabbari-Farouji, J. Chem. Phys. \textbf{159}, 224903 (2023).

\bibitem{ViscoBath}
H. Vandebroek and C. Vanderzande, Phys. Rev. E \textbf{92}, 060601 (2015).

\bibitem{EnzymActive}
S. Put, T. Sakaue and C. Vanderzande, Phys. Rev. E \textbf{99}, 032421 (2019).

\bibitem{Duke_PRL}
T. Duke, T. E. Holy and S. Leibler, Phys. Rev. Lett. \textbf{74}, 330 (1995).

\bibitem{Bourdieu_PRE}
L. Bourdieu, M. O. Magnasco, D. A. Winkelmann and A. Libchaber, Phys. Rev. E \textbf{52}, 6573 (1995).

\bibitem{Bourdieu_PRL}
L. Bourdieu, T. Duke, M. B. Elowitz, D. A. Winkelmann, S. Leibler and A. Libchaber, Phys. Rev. lett.  \textbf{75}, 176 (1995).

\bibitem{Grill_Nano_Lett}
C. Reuther, M. Mittasch, S. R. Naganathan, S. W. Grill and S. Diez, Nano Lett. \textbf{17}, 5699 (2017).

\bibitem{Tas_Nano_Lett}
R. P. Tas, C.-Y. Chen, E. A. Katrukha, M. Vleugel, M. Kok, M. Dogterom, A. Akhmanova and L. C. Kapitein,   \textbf{18}, 7524 (2018).

\bibitem{Farkas_book}
Z. Farkas, I. Derenyi and T. Vicsek, {\it Structure and Dynamics of Confined Polymers} (Springer, Netherlands, 2002).




\bibitem{wca}
J. D. Weeks, D. Chandler, and H. C. Andersen, J. Chem. Phys. 
\textbf{54}, 5237 (1971).

\bibitem{Amir}
A. Rezaie-Dereshgi, H. Khalilian and J. Sarabadani, J. Phys.: Condens. Matter, \textbf{35}, 355101 (2023).

\bibitem{lammps}
S. Plimpton, J. Comp. Phys. \textbf{117}, 1-19 (1995), http://lammps.sandia.gov.

\bibitem{turn_handbook}
S. G. Krantz, {\it Handbook of Complex Variables}, (Birkhäuser,
Boston, MA, 1999).




\bibitem{Panja_J_Stat_Mech}
D. Panja, J. Stat. Mech.: Theory and Experiment \textbf{10}, 1742 (2010).

\bibitem{Vandebroek_J_Stat_Phys}
H. Vandebroek and C. Vanderzande, J. stat. Phys. \textbf{167}, 14 (2017).

\bibitem{Sakaue_PRE_2013}
T. Sakaue, Phys. Rev. E \textbf{87}, 040601 (2013).

\bibitem{Saito_PRE_2015}
T. Saito and T. Sakaue, Phys. Rev. E \textbf{92}, 012601 (2015).

\bibitem{Shinkai}
S. Shinkai, S. Onami and T. Miyaguchi, arXiv:2405.04019v1.

\bibitem{Polymer_folding_PRE_2102}
J.-C. Walter, A. Ferrantini, E. Carlon and C. Vanderzande, Phys. Rev. E \textbf{85}, 031120 (2012).

\bibitem{Polymer_folding_PRL_2104}
R. Frederickx, T. in’t Veld and E. Carlon, Phys. Rev. Lett. \textbf{112}, 198102 (2014).

\bibitem{Soft_Matter_2017}
T. Sakaue, J.-C. Walter, E. Carlon and C. Vanderzande, Soft Matter \textbf{13}, 3174 (2017).


\bibitem{Poly_trans_J_Cond_Matt_2013}
D. Panja, G. T. Barkema and A. B. Kolomeisky, J. Phys.: Condens. Matter \textbf{25}, 413101 (2013).

\bibitem{Poly_trans_Soft_Matter_2014}
V. V. Palyulin, T. Ala-Nissila and R. Metzler, Soft Matter \textbf{10}, 9016 (2014).

\bibitem{Poly_trans_Polymers_2016}
T. Sakaue, Polymers \textbf{8}, 424 (2016).

\bibitem{Sakaue_PRE_2007}
T. Sakaue, Phys. Rev. E \textbf{76}, 021803 (2007).

\bibitem{Rowghanian_2011}
P. Rowghanian and A. Y. Grosberg, J. Phys. Chem. B \textbf{115}, 14127 (2011).

\bibitem{Timo_PRE_2012}
T. Ikonen, A. Bhattacharya, T. Ala-Nissila and W. Sung, Phys. Rev. E \textbf{85}, 051803 (2012).

\bibitem{Jalal_JCP_2014}
J. Sarabadani, T. Ikonen and T. Ala-Nissila, J. Chem. Phys. \textbf{141}, 214907 (2014).

\bibitem{Jalal_JCP_2015}
J. Sarabadani, T. Ikonen and T. Ala-Nissila, J. Chem. Phys. \textbf{143}, 074905 (2015).

\bibitem{Jalal_EPL_2017}
J. Sarabadani, B. Ghosh, S. Chaudhury and T. Ala-Nissila, Europhys. Lett. \textbf{120}, 38004 (2017).

\bibitem{Jalal_Sci_Rep_2017}
J. Sarabadani, T. Ikonen, H. M$\ddot{\textrm{o}}$kk$\ddot{\textrm{o}}$nen, T. Ala-Nissila, S. Carson and M. Wanunu, Sci. Rep. \textbf{7}, 7423 (2017).

\bibitem{Jalal_Cond_Matt_2018}
J. Sarabadani and T. Ala-Nissila, J. Phys.: Condens. Matter \textbf{30}, 274002 (2018).

\bibitem{Jalal_Cond_Matt_2020}
B. Ghosh, J. Sarabadani, S. Chaudhury and T. Ala-Nissila, J. Phys.: Condens. Matter \textbf{33}, 015101 (2020).

\bibitem{Jalal_PRR_2021}
H. Khalilian, J. Sarabadani and T. Ala-Nissila, Phys. Rev. Research \textbf{3}, 013080 (2021).

\bibitem{Jalal_PRR_2023}
H. Khalilian, J. Sarabadani and T. Ala-Nissila, Phys. Rev. Research \textbf{5}, 023107 (2023).


\bibitem{strogatz}
Steven H. Strogatz, {\it Nonlinear Dynamics and Chaos}, (Perseus Books Publishing, L.L.c., 1994).

\bibitem{van_Kampen}
N. G. van Kampen, {\it Stochastic Processes in Physics and Chemistry}, (North Holand, The Netherlands, 2007).

\bibitem{Reif}
F. Reif, {\it Fundamentals of Statistical and Thermal Physics}, (Waveland Press Inc., The United States of America, 2009).




\bibitem{Sokolov_PRE_2012}
F. Thiel, L. Schimansky-Geier and I. M. Sokolov, Phys. Rev. E \textbf{86}, 021117 (2012).

\bibitem{OptimalLangevin}
C. Micheletti, G. Bussi and A. Laio, J. Chem. Phys. \textbf{129}, 074105 (2008).



\bibitem{rubin_stein}
M. Rubinstein and R. Colby, {\it Polymer Physics}, (Oxford University Press, New York, 2003).






\end{thebibliography}

\end{document}